\documentclass[aps,fleqn,usenatbib, twocolumn, superscriptaddress]{revtex4}
\usepackage{amsmath}%
\usepackage{amsfonts}%
\usepackage{amssymb}%
\usepackage{graphicx}
\usepackage{color}

\newcommand\dderiv{{\mathrm{d}}}

\begin{document}
\title{Microwave Background Polarization as a Probe of Large-Angle Correlations}

\author{Amanda Yoho} 
\affiliation{CERCA/ISO, Department of Physics, Case Western Reserve University,
10900 Euclid Avenue, Cleveland, OH 44106-7079, USA}
\author{Simone Aiola}
\affiliation{Department of Physics and Astronomy, University of Pittsburgh, Pittsburgh, PA 15260 USA} 
\affiliation{Pittsburgh Particle Physics, Astrophysics, and Cosmology Center (PITT-PACC), Pittsburgh PA 15260}
\author{Craig J. Copi}
\affiliation{CERCA/ISO, Department of Physics, Case Western Reserve University,
10900 Euclid Avenue, Cleveland, OH 44106-7079, USA}
\author{Arthur Kosowsky}
\affiliation{Department of Physics and Astronomy, University of Pittsburgh, Pittsburgh, PA 15260 USA} 
\affiliation{Pittsburgh Particle Physics, Astrophysics, and Cosmology Center (PITT-PACC), Pittsburgh PA 15260}
\author{Glenn ~D.~Starkman}
\affiliation{CERCA/ISO, Department of Physics, Case Western Reserve University,
10900 Euclid Avenue, Cleveland, OH 44106-7079, USA}

\begin{abstract}
Two-point correlation functions of cosmic microwave background
polarization provide a physically independent probe of the surprising
suppression of correlations in the cosmic microwave background
temperature anisotropies at large angular scales. We investigate correlation
functions constructed from both the Q and U Stokes parameters and from
the E and B polarization components. The dominant contribution to
these correlation functions comes from local physical effects at the
last scattering surface or from the epoch of reionization at high
redshift, so all should be suppressed if the temperature suppression
is due to an underlying lack of correlations in the cosmological metric
perturbations larger than a given scale. We evaluate the correlation
functions for the standard $\Lambda$CDM cosmology constrained by the observed
temperature correlation function, and compute statistics
characterizing their suppression on large angular scales. Future
full-sky polarization maps with minimal systematic errors on large
angular scales will provide strong tests of whether the observed
temperature correlation function is a statistical fluke or reflects a
fundamental shortcoming of the standard cosmological model. 

\end{abstract}
\maketitle

%%%%%%%%%%%%%%%%%%%%%%%%
\section{Introduction}
%%%%%%%%%%%%%%%%%%%%%%%%

Two seasons of observational data from the Planck satellite have given us 
the most precise measurement of temperature fluctuations in the Cosmic Microwave Background
on the full sky to date~\cite{Ade:2013zuv, Adam:2015rua, Planck:2015xua}. These observations appear to fit well within the standard picture of our Universe --
Lambda Cold Dark Matter ($\Lambda$CDM). It did, however confirm several anomalous features
in the temperature fluctuations~\cite{Ade:2013nlj}, which had first been hinted at with the COBE-DMR satellite~\cite{Bennett:1996ce} and were
later highlighted in the WMAP data releases~\cite{Spergel:2003cb}. These anomalies exist overwhelmingly at the largest 
scales of the temperature power spectrum, $C^{TT}_{\ell}$, with several interesting features appearing
 at multipoles $\ell\lesssim 30$. One feature, the lack of two-point correlation at angular separations of $60^{\circ}$ and 
 above, has garnered much attention recently~\cite{Copi:2010na, Copi:2008hw}.
With decades of temperature measurements in hand, we know that this lack of correlation occurs only $0.03 - 0.1$ per cent
of the time in $\Lambda$CDM realizations.

These large scales are also where cosmic variance, rather than statistical errors, is the limiting factor in our ability to compare the observed
value of $C_{\ell}^{TT}$ to its theoretical value.
 This means that additional measurements of the temperature fluctuations
will not help us make more definitive statements about the nature of the lack of correlation, and whether 
it is a statistical fluke within our cosmological model or due to unknown physics. 
  Work has been done recently
to quantify the viability of using cross correlations of temperature with E-mode polarization~\cite{Copi:2013zja} and 
the lensing potential $\varphi$~\cite{Yoho:2013tta} to test this ``fluke hypothesis." 

Correlations of CMB polarization itself, outside of just cross correlations with the temperature observations,
are a natural next step in determining the nature of the lack of temperature correlation seen at large angles. 
A feature that is required for a real-space correlation function is for the field to be calculated
using only local operators on directly observed $Q$ and $U$ polarization maps. The very nature
of a correlation function that has a clearly defined physical interpretation depends on points on the sky 
being determined independently of each other (i.e. locally).

To accomplish this, we calculate two sets of polarization correlation functions:
$Q$ and $U$ auto-correlations along with $\hat{E}({\bf\hat{n}})$ and $\hat{B}({\bf\hat{n}})$ auto-correlations.
These have a number of properties that make them unique tests of large-angle 
correlation suppression, such as
 contributions from the reionization bump that appear in polarization power spectra at $\ell\lesssim 10$
 that dominate the large-angle $Q$ and $U$ functions. 
The local E- and B-mode correlations are instead dominated by large multipoles at large angles, and have
small contributions from reionization which makes them a cleaner test
of physics at the last scattering surface. 
In this work we present the local $C^{\hat{E}\hat{E}}(\theta)$ and $C^{\hat{B}\hat{B}}(\theta)$, along with 
$C^{QQ}(\theta)$ and $C^{UU}(\theta)$, and show distributions for the corresponding $S_{1/2}$ 
statistic for each. These results are drawn using constrained temperature realizations, meaning they are consistent with the observed power
 spectrum within instrumental errors and have a cut-sky $S_{1/2}$ at least as small as our cut-sky measurement.

This paper is organized as follows: in Section~\ref{background} we present the theoretical background for $C(\theta)$ and 
a commonly discussed statistic $S_{1/2}$, in Section~\ref{errors} we discuss our calculation of the error based on
next-generation satellite specifications as well as the lowest possible expected instrument-limited value of $S_{1/2}$, in Section~\ref{localB} we present the local E- and B-mode correlation functions,
in Section~\ref{QQUU} we show auto-correlation functions for Q and U Stokes parameters, and in Section~\ref{conclusions}
we present our conclusions and discuss possibilities for future work.

%%%%%%%%%%%%%%%%%%%%%%%%
\section{Background}\label{background}
%%%%%%%%%%%%%%%%%%%%%%%%
\subsection{Temperature Correlation Function and Statistics}
The information contained in CMB temperature fluctuations is often represented in harmonic space
by decomposing them in terms of spherical harmonics and their coefficients,
\begin{equation}
\frac{\Delta T({\bf \hat{n}})}{T_{o}}\equiv\Theta({\bf \hat{n}}) = \sum_{\ell, m}a^{T}_{\ell m}Y_{\ell m}({\bf \hat{n}}),
\end{equation}
with the temperature power spectrum being constructed from the $a_{\ell m}$ coefficients:
\begin{equation}
\langle a^{T}_{\ell m} a^{T*}_{\ell^{\prime}m^{\prime}} \rangle = \delta_{\ell\ell^{\prime}}\delta_{mm^{\prime}}C^{TT}_{\ell}
\end{equation}

In real space, the CMB temperature fluctuations, $\Theta({\bf \hat{n}})$, can be represented as a two-point correlation function
averaged over the sky at different angular separations:
\begin{equation}\label{tcorr}
C^{TT}(\theta)=\overline{\Theta({\bf \hat{n}_1})\Theta({\bf \hat{n}_2})}   \quad  {\mathrm{with}} \quad 
{\bf \hat{n}_1}\cdot{\bf \hat{n}_2}=\cos\theta.
\end{equation}
This is an estimator of the quantity 
$C^{TT}(\theta)=\langle\Theta({\bf \hat{n}_1})\Theta({\bf \hat{n}_2})\rangle$, where the angle brackets
represent an ensemble average.
The sky average over the angular separation can be expanded in a Legendre series,
\begin{equation}\label{theta_ell}
C^{TT}(\theta)=\sum_{\ell}\frac{2\ell+1}{4\pi}\,C_{\ell}^{TT}P_{\ell}(\cos\theta),
\end{equation}
where the $C_{\ell}$ on the right-hand side of Eq.~(\ref{theta_ell}) are the pseudo-$C_{\ell}$
temperature power spectrum values.

The $S_{1/2}$ statistic was defined by the WMAP team to quantify the lack of angular correlation seen in 
temperature maps~\cite{Spergel:2003cb}: 

\begin{equation}\label{Shalf}
  S_{1/2}^{TT} \equiv \int_{-1}^{1/2}\dderiv(\cos\theta) [C^{TT}(\theta)]^2.
\end{equation}
The expression for $S_{1/2}$ can be written conveniently in terms of the temperature
power spectrum and a coupling matrix $I_{\ell\ell^{\prime}}$,
\begin{equation}\label{Illp}
  S_{1/2}^{TT} = \sum_{\ell=2}^{\ell_{\mathrm max}} C_{\ell}^{TT} {I}_{\ell\ell'} C_{\ell'}^{TT}.
\end{equation} 
 A full 
expression of the $I_{\ell \ell^{\prime}}$ matrix can be found in Appendix B of \citep{Copi:2008hw}.
The $C_{\ell}$ fall sharply and higher order modes have a 
negligable contribution to the statistic, so choice of an appropriately
large value of $\ell_{\mathrm{max}}$ in Eq.~(\ref{Shalf}) will ensure that the result is not affected
by including additional higher-$\ell$ terms.

\subsection{Stokes $Q$ and $U$ Correlation Functions and Statistics}
Linear polarization is typically described by two quantities: the $Q$ and $U$ Stokes parameters in real space, and
E-modes and B-modes in harmonic space. 
In real space, 
$C^{QQ}(\theta) =\langle{Q_{r}({\bf \hat{n}_1})Q_{r}({\bf \hat{n}_2})}\rangle$
and $C^{UU}(\theta) =\langle U_{r}({\bf \hat{n}_1})U_{r}({\bf \hat{n}_2})\rangle$ are the $Q$ and $U$ correlation functions,
where $Q_{r}({\bf \hat{n}})$ and $U_{r}({\bf \hat{n}})$ are the Stokes parameters
defined with respect to the great arc connecting ${\bf \hat{n}_1}$ and ${\bf \hat{n}_2}$~\cite{Kamionkowski:1996zd}.
$Q({\bf \hat{n}})$ and $U({\bf \hat{n}})$ fields on the sphere are defined such that they are 
connected by a great arc of constant $\phi$.
In practice, the correlation 
functions are calculated as an average over pixels separated by an angle $\theta$: 
\begin{eqnarray}\label{QUcorr}
C^{QQ}(\theta) &=\overline{Q_{r}({\bf \hat{n}_1})Q_{r}({\bf \hat{n}_2})},\notag \\
C^{UU}(\theta) &=\overline{U_{r}({\bf \hat{n}_1})U_{r}({\bf \hat{n}_2})}.
\end{eqnarray}

The decomposition of polarization into spin-2 spherical
harmonics is done with a linear combination of the Stokes parameters,
\begin{equation}\label{QUdef}
\left(Q({\bf \hat{n}}) \pm \mathrm{i} U({\bf \hat{n}}) \right) = \sum_{\ell m} \;_{\pm2}a^{P}_{\ell m} \;_{\pm2}Y_{\ell m}({\bf \hat{n}}).
\end{equation}
The standard E- and B-mode coefficients are combinations of the spin-2 harmonic coefficients,
\begin{eqnarray}\label{eq:almEB}
a_{\ell m}^{B} & = \frac{\mathrm{i}}{2} \left[ _{2}a^{P}_{\ell m} - \;_{-2}a^{P}_{\ell m}\right] \notag \\
a_{\ell m}^{E} & = -\frac{1}{2} \left[ _{2}a^{P}_{\ell m} + \;_{-2}a^{P}_{\ell m}\right],
\end{eqnarray}
and the E- and B-mode power spectra are defined as
\begin{eqnarray}
\langle a^{E}_{\ell m} {a^{E}}^{*}_{\ell^{\prime}m^{\prime}} \rangle &= \delta_{\ell\ell^{\prime}}\delta_{mm^{\prime}}C^{EE}_{\ell}\notag\\
\langle a^{B}_{\ell m} {a^{B}}^{*}_{\ell^{\prime}m^{\prime}} \rangle &= \delta_{\ell\ell^{\prime}}\delta_{mm^{\prime}}C^{BB}_{\ell}.
\end{eqnarray}

Using these equations, we can construct $C^{QQ}(\theta)$ and $C^{UU}(\theta)$ 
from  $C_{\ell}^{BB}$ and  $C_{\ell}^{EE}$~\cite{Kamionkowski:1996ks}:
\begin{align}\label{QU}
C^{QQ}(\theta) &= -\sum_{\ell} \frac{2\ell + 1}{4\pi} \left( \frac{2(\ell -2)!}{(\ell +2)!}\right)\times\notag\\
 &\left[ C_{\ell}^{EE} G_{\ell 2}^{+}(\cos\theta)+C_{\ell}^{BB} G_{\ell 2}^{-}(\cos\theta) \right]\notag\\
C^{UU}(\theta) &= -\sum_{\ell} \frac{2\ell + 1}{4\pi} \left( \frac{2(\ell -2)!}{(\ell +2)!}\right)\times\notag\\
&\left[C_{\ell}^{EE} G_{\ell 2}^{-}(\cos\theta) + C_{\ell}^{BB} G_{\ell 2}^{+}(\cos\theta)\right],
\end{align}
where
\begin{align}
G_{\ell m}^{+}(\cos\theta) &= - \left( \frac{\ell-m^2}{\sin^2\theta}+\frac{\ell(\ell-1)}{2}P_{\ell}^{m}(\cos\theta) \right)  \notag\\
                         &+  (\ell+m)\frac{\cos\theta}{\sin^2\theta}P_{\ell-1}^{m}(\cos\theta),\notag \\
G_{\ell m}^{-}(\cos\theta) &= \frac{m}{\sin^2\theta}\left( (\ell-1)\cos\theta P_{\ell}^{m}(\cos\theta)\right. \notag\\
                         &\left. -(\ell+m)P_{\ell-1}^{m}(\cos\theta) \right).
\end{align}

The $G_{\ell m}^{\pm}(\cos\theta)$ are complicated functions of Legendre polynomials, so the calculation of 
$S_{1/2}^{QQ}$ and $S_{1/2}^{UU}$ is not a straightforward analog to Eq.~(\ref{Illp}). Instead, there will be three terms:
\begin{align}\label{IQU}
  S_{1/2}^{QQ} = & \sum_{\ell=2}^{\ell_{\mathrm max}} C_{\ell}^{EE} {I}^{(1)}_{\ell\ell'} C_{\ell'}^{EE}  
  +  C_{\ell}^{BB} {I}^{(3)}_{\ell\ell'} C_{\ell'}^{BB}\notag\\
  & + 2 C_{\ell}^{EE} {I}^{(2)}_{\ell\ell'} C_{\ell'}^{BB},
\end{align} 
where for $S_{1/2}^{UU}$ the ${I}^{(1)}_{\ell\ell'}$ and ${I}^{(3)}_{\ell\ell'}$ are swapped. Full details
of calculating the ${I}^{(i)}_{\ell\ell'}$ matrices is outlined in Appendix~\ref{dmat}.

\subsection{E- and B-mode Correlation Functions and Statistics}

The local correlation functions on the sky of the E- and B-modes are defined as
\begin{align}\label{polcorr}
C^{\hat{B}\hat{B}}(\theta)& =  \langle \hat{B}({\bf \hat{n}_1}) \hat{B}({\bf \hat{n}_2})\rangle \notag\\
C^{\hat{E}\hat{E}}(\theta)& =  \langle \hat{E}({\bf \hat{n}_1}) \hat{E}({\bf \hat{n}_2})\rangle.
\end{align}
The $\hat{E}({\bf \hat{n}})$ and $\hat{B}({\bf \hat{n}})$ functions can be calculated from
the observable Q and U fields using {\it local} spin raising and lowering operators $\bar{\eth}$ and $\eth$~\cite{Zaldarriaga:1996xe}:
\begin{align}\label{EBdef}
\hat{B}({\bf \hat{n}}) & = \frac{-\mathrm{i}}{2}\left[\bar{\eth}^2(Q({\bf \hat{n}})+\mathrm{i} U({\bf \hat{n}})) - \eth^2(Q({\bf \hat{n}})-\mathrm{i} U({\bf \hat{n}}))\right]\notag\\
\hat{E}({\bf \hat{n}}) & =\frac{1}{2}\left[\bar{\eth}^2(Q({\bf \hat{n}})+\mathrm{i} U({\bf \hat{n}})) + \eth^2(Q({\bf \hat{n}})-\mathrm{i} U({\bf \hat{n}}))\right], 
\end{align}
where 
\begin{align}\label{eth}
\eth & = -(\sin\theta)\left[ \frac{\partial}{\partial\theta} + \left(\frac{\mathrm{i}}{\sin\theta}\right)\frac{\partial}{\partial\phi}\right] (\sin\theta)^{-1}, \notag\\
\bar{\eth} & = -(\sin\theta)^{-1}\left[ \frac{\partial}{\partial\theta} - \left(\frac{\mathrm{i}}{\sin\theta}\right)\frac{\partial}{\partial\phi}\right] (\sin\theta)
\end{align}
in real space, and in harmonic space,
\begin{align}
\eth\;_{s}Y_{\ell m} & = \sqrt{({\ell-s})(\ell+s+1)} \;_{s+1}Y_{\ell m}, \notag\\
\bar{\eth}\;_{s}Y_{\ell m} & = -\sqrt{({\ell+s})(\ell-s+1)} \;_{s-1}Y_{\ell m}.
\end{align}

In terms of spherical harmonics and coefficients, $\hat{E}({\bf \hat{n}})$ and $\hat{B}({\bf \hat{n}})$ are~\cite{Kamionkowski:1996ks, Zaldarriaga:1996xe}:
\begin{align}\label{poldef}
\hat{B}({\bf \hat{n}}) & = \sum_{\ell m} \sqrt{\frac{(\ell+2)!}{(\ell -2)!}}\; a_{\ell m}^{B}Y_{\ell m}({\bf \hat{n}})\notag\\
\hat{E}({\bf \hat{n}}) & = \sum_{\ell m} \sqrt{\frac{(\ell+2)!}{(\ell -2)!}}\; a_{\ell m}^{E}Y_{\ell m}({\bf \hat{n}}).
\end{align}
The prefactor under the square root is proportional to $\ell^4$, and is a direct consequence of using the local operators 
on the $Q$ and $U$ maps. 

Real-space fields of E- and B-modes are occasionally presented as spin-zero quantities~\cite{Baumann:2009mq}, 
\begin{align}\label{eq:polspinzero}
E({\bf \hat{n}}) & \equiv \sum_{\ell m} a_{\ell m}^{E}Y_{\ell m}({\bf \hat{n}}),\notag \\
B({\bf \hat{n}}) & \equiv \sum_{\ell m} a_{\ell m}^{B}Y_{\ell m}({\bf \hat{n}}).
\end{align}
The fields in Eq.~(\ref{eq:polspinzero}) {\it cannot} be constructed from real-space maps only, unlike Eq.~(\ref{eq:polspinzero}), and require map filtering
in harmonic space to separate the E- and B-modes. 
Because polarization is inherently a spin-2 quantity and an integral over the full sky is required 
to extract the $a_{\ell m}^{E}$ and $a_{\ell m}^{B}$ coefficients from Eqs.~\ref{QUdef} and~\ref{eq:almEB}, the
$E({\bf \hat{n}})$ and $B({\bf \hat{n}})$ are {\it non-local}. The non-local definitions of $E({\bf \hat{n}})$ and $B({\bf \hat{n}})$ require
information from the full sky to separate the E- and B- modes from observed Q and U polarization maps in any given 
pixel. For this reason,
non-local definitions cannot be used when talking about real-space correlation functions, since the physical interpretation
of a correlation at one particular point on the sky ${\bf \hat{n}_1}$ with another particular point on the sky ${\bf \hat{n}_2}$
becomes ambiguous.

The expression for the two point function in terms of the local fields is
\begin{equation}\label{Cxy} 
C^{\hat{E}\hat{E}}(\theta)=\sum_{\ell}\frac{2\ell+1}{4\pi}\left(\frac{(\ell+2)!}{(\ell -2)!}\right)\;C^{EE}_{\ell}P_{\ell}(\cos\theta),
\end{equation}
and the same for the local $\hat{B}$ correlation when substuting in $C_{\ell}^{BB}$.
 This form of the correlation function leads to some interesting conclusions, namely
that the traditional mode of thinking that $\theta\sim\frac{1}{\ell}$ is not applicable. This intuition was due 
directly to the fact that $C_{\ell}^{TT}$ falls off as $1/\ell^2$ and the prefactor in the sum for the $TT$
correlation function in Eq.~(\ref{theta_ell})
only scales like $\ell$, leaving the sum dominated by terms less than an $\ell_{\mathrm{max}}= 30$. This does not hold for correlation functions of the $\hat{E}({\bf \hat{n}})$ and $\hat{B}({\bf \hat{n}})$ functions defined in
Eq.~(\ref{poldef}),
and it should be clear that higher $\ell$ modes
will contribute to the large-angle piece of the correlation functions. This feature was also discussed in~\cite{Baumann:2009mq}, where they were focused on small-angle correlation functions of local E- and B-modes.

The expressions for $S_{1/2}^{\hat{E}\hat{E}}$ and $S_{1/2}^{\hat{B}\hat{B}}$ are similar to Eq.~(\ref{Illp}):
\begin{equation}\label{IEB}
  S_{1/2}^{XX} = \sum_{\ell, \ell^{\prime}=2}^{\ell_{\mathrm max}, \ell^{\prime}_{\mathrm max}} \left(\frac{(\ell+2)!}{(\ell -2)!}\right)C_{\ell}^{XX} 
  I_{\ell\ell'} \left(\frac{(\ell'+2)!}{(\ell' -2)!}\right)C_{\ell'}^{XX}.
\end{equation}

We have chosen to calculate the $S_{1/2}$ statistic, rather than generalizing to a statistic at another angle, because
effects that contribute to polarization inside the surface of last scattering (namely reionization) are at a sufficiently
high redshift that they do not significantly change the relevant angle where suppression is expected to appear.

%%%%%%%%%%%%%%%%%%%%%%
\section{Error limits on measuring a suppressed $C(\theta)$ for future CMB polarization experiments}\label{errors}
%%%%%%%%%%%%%%%%%%%%%%

The error in $C_{\ell}$ for a next-generation full-sky CMB satellite can be determined
using the relation 

\begin{equation}\label{pixie}
\Delta C_{\ell} = \sqrt{\frac{2}{2\ell + 1}}\left( C_{\ell} + \frac{e^{\ell^{2}\sigma_{b}^{2}}\sigma^2}{4\pi}\right),
\end{equation}
where $\sigma$ is the pixel error estimate in $\mu\mathrm{K}-\mathrm{arcmin}$~\cite{Knox:1995dq}. Values for the
pixel error estimates for future
surveys are shown in Table~\ref{params}~\cite{Ade:2013ktc, Kogut:2011xw, Andre:2013afa}.  

\begin{center}
\begin{table}[hc]
\begin{tabular}{ccc}
Experiment & $\;\;\;\;\;\;\;\sigma_{P}$ [$\mu\mathrm{K}\;\mathrm{arcmin}$] & $\theta_{\mathrm{FWHM}}$ [arcmin]\\
\hline 
Planck & $120$ & 5 \\ 
\hline 
PIXIE & $3.78$ & 54 \\ 
\hline 
PRISM & $3.4$ & 2 \\ 
\hline 
\end{tabular} 
\caption{Polarization sensitivities that reflect the
actual Planck sensitivity in CMB channels, and
the design sensitivity for two satellite proposals.}\label{params}
\end{table}
\end{center}

To find the corresponding error band in $C(\theta)$, we create $10^5$ realizations of the $C_{\ell}^{BB}$
spectrum assuming chi-squared distribution with variance including instrumental error based on the values 
in Table~\ref{params}. Constrained realizations 
of $C_{\ell}^{EE}$ are generated by drawing $a_{\ell m}^E$ coefficients using instrument noise and assuming they are coupled 
to constrained realizations of $a_{\ell m}^T$.
 
The constrained temperature harmonic coefficients are drawn such that they produce $S_{1/2}$ 
values that are consistent with calculations from data and have a spectrum which matches observations (the full procedure
for making constrained realizations is outlined in~\cite{Copi:2013zja}). The errors to the mean correlation
function values are determined based on the $68\%$ confidence levels (C.L) for the realizations.
Cosmic variance dominates the error bars on the E- and B-mode power spectra through the reionization
 bump ($\ell\lesssim 10$) and instrumental error from beam size dominates around $\ell\sim 45$ for $r=0.1$.
 
The instrumental error enforces a limit on the smallest possible value for the expectation $\langle S_{1/2} \rangle$,
even if the correlation function is completely suppressed.
If we assume that the correlation functions defined in Eqs.~\ref{QU} and~\ref{Cxy} are noise-free and 
identically zero above 60 degrees, then the corresponding sums over the power spectra and their coefficients must be zero
for all $P_{\ell}(\cos\theta<1/2)$. 
For both sets of correlation functions, this makes 
$S_{1/2}$ for $Q$, $U$, $\hat{E}$ or $\hat{B}$
\begin{equation}\label{exIQU}
  S_{1/2} = \int_{-1}^{1/2}[\delta C^{XX}(\theta)]^2\;\; \mathrm{d}\cos\theta.
\end{equation} 

In real-space, for $Q$
\begin{equation}\label{exIQU}
\delta C^{QQ}(\theta) = \frac{\sigma_{P}}{\sqrt{2\; N_{\mathrm{pairs}}}}Q_{\mathrm{rms}},
\end{equation}
where $N_{\mathrm{pairs}}$ is the number of pixel pairs separated by $\theta$ 
and $Q_{\mathrm{rms}}$ is the root mean square value of the field.
The integral is trivial since the only $\theta$ dependence appears
in the expression for $N_{\mathrm{pairs}}$:
\begin{equation}\label{Npairs}
N_{\mathrm{pairs}} = \frac{1}{2} N_{\mathrm{pix}}^{3/2}\pi^{1/2}\sin\theta.
\end{equation}
The zero true-sky value of $S_{1/2}$ is
\begin{equation}
S_{1/2}^{QQ} = \frac{3\;\sigma_{P}^2\;Q_{\mathrm{rms}}^2}{\;2N_{\mathrm{pix}}^{3/2}\pi^{1/2}}.
\end{equation}
This result is the same for the $U$ field, with $U_{\mathrm{rms}}$ substituted 
for $Q_{\mathrm{rms}}$.

For the E-mode statistics, it is easier to calculate $\delta C(\theta)$ in $\ell$-space:  
\begin{align}\label{exIEB}
  \delta &C^{\hat{E}\hat{E}}(\theta) = \frac{1}{\sqrt{8\pi N_{\mathrm{pairs}}}} \times \notag\\
  & \sqrt{\sum_{\ell \ell^{'}}
  \left(\frac{(\ell+2)!}{(\ell-2)!}\right)^2(2\ell + 1)(2\ell^{'}+1)C^{EE}_{\ell}N^{EE}_{\ell^{'}}}.
\end{align} 

This leads to
\begin{align}
S_{1/2}^{\hat{E}\hat{E}} = &&\frac{3}{8(N_{\mathrm{pix}}\pi)^{3/2}}\sum_{\ell\ell^{\prime}}C_{\ell}^{EE}(2\ell+1)\times\notag\\
& & \left(\frac{(\ell+2)!}{(\ell-2)!}\right)^2
N^{EE}_{\ell^{\prime}}(2\ell^{\prime}+1),
\end{align}
with the same result for $\hat{B}$ when $C_{\ell}^{BB}$ is substituted for 
$C_{\ell}^{EE}$, and using $N^{BB}_{\ell^{\prime}} = N^{EE}_{\ell^{\prime}}$.

In the near term, Planck will weigh in with its upcoming release of polarization data. We do not yet know the exact
noise spectra for their $EE$ and $BB$ observations, but we can make an estimate of the expected $S_{1/2}$ values assuming
$\sigma_{\mathrm{pol}} = \sqrt{2}\;\sigma_T$ and using $\sigma_{T} = 85\mu\mathrm{K}-\mathrm{arcmin}$ from~\cite{Ade:2013ktc}. Table~\ref{params} outlines error estimates used for Planck in addition to PIXIE~\cite{Kogut:2011xw} and 
PRISM~\cite{Andre:2013afa}, and Table~\ref{tab:S12_exp} presents all values of the $S_{1/2}$ statistic 
that results from assuming there is zero true correlation at the last scattering surface for each experiment. 
These values show that, when compared to the $\Lambda$CDM prediction of $S_{1/2}$,
 pixel noise is not a significant source of error to quantifying suppression
to the correlation functions in polarization. Systematic errors may bias measurements of $S_{1/2}$,
but we will not consider these here as any unresolved systematic would only serve
to {\it increase} the value of $S_{1/2}$. Currently, no full-sky
polarization maps are reliable enough to measure the
large-angle polarization functions computed here.

\begin{center}
\begin{table}[hc]
\begin{tabular}{cccc}
Experiment & $\;\;\;\;\;\;\;QQ/UU[\mu\mathrm{K}^4]$ & $\hat{E}\hat{E}[\mu\mathrm{K}^4]$ &$ \hat{B}\hat{B}[\mu\mathrm{K}^4]$  \\
\hline 
Planck & $1.75\times 10^{-6} $ & $0.314$ & $0.013$ \\ 
\hline 
PIXIE & $1.73\times 10^{-9}$ & $3.10\times 10^{-4}$ & $1.31\times 10^{-5}$ \\ 
\hline 
PRISM & $1.40\times 10^{-9}$ & $2.51\times 10^{-4}$ & $1.06\times 10^{-5}$ \\ 
\hline 
\end{tabular} 
\caption{Expected values of $S_{1/2}$ statistic from a toy-model map with pixel noise using
sensitivites from Table~\ref{params}
and assuming complete suppression of the true correlation function
 for $Q$, $U$, $\hat{E}$, $\hat{B}$. These
estimates account for sensitivities for future CMB polarization satellites.}\label{tab:S12_exp}
\end{table}
\end{center}

\section{Local $\hat{B}({\bf\hat{n}})$ and $\hat{E}({\bf\hat{n}})$ Correlation Functions}\label{localB}

\begin{figure}
\includegraphics[scale=.6]{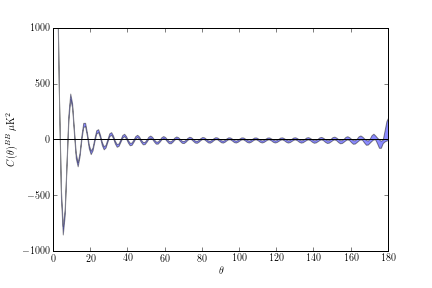}
\caption{Angular correlation function of local B-modes $r=0.1$ with $\sigma_{\text{beam}}=2.7^{\circ}$ smoothing. The blue shaded 
region corresponds to $68\%$ C.L. errors, which includes instrumental noise for a future generation PIXIE-like 
experiment and cosmic variance using Eq.~(\ref{pixie}).}\label{CthetaBB_r01}
\end{figure}

\begin{figure}
\includegraphics[scale=.6]{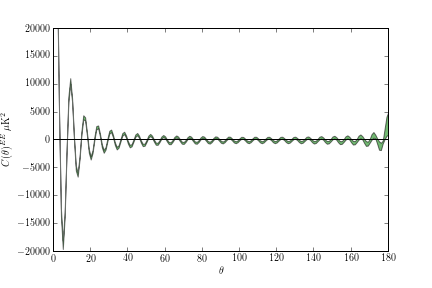}
\caption{Angular correlation function of constrained local E-modes $r=0.1$ with $\sigma_{\text{beam}}=2.7^{\circ}$ 
smoothing. The green shaded 
region corresponds to $68\%$ C.L. errors, which includes instrumental noise for a future generation PIXIE-like 
experiment and cosmic variance using Eq.~(\ref{pixie}).}\label{CthetaEE_r01}
\end{figure}

\begin{figure}
\includegraphics[scale=.6]{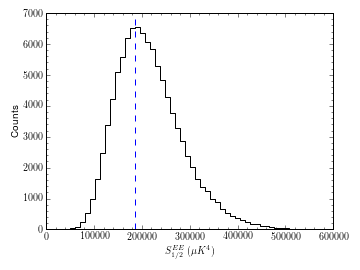}
\caption{$S_{1/2}$ statistic distribution for the angular correlation function of E-modes $r=0.1$ with $\sigma_{\text{beam}}=2.7^{\circ}$ radian smoothing. The blue dashed line marks the $\Lambda$CDM prediction for the ensemble average.}\label{S12EE_r01}
\end{figure}

\begin{figure}
\includegraphics[scale=.6]{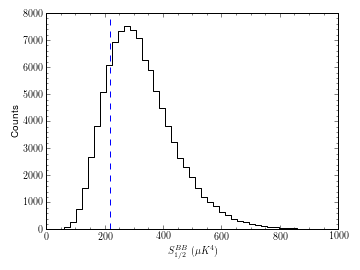}
\caption{$S_{1/2}$ statistic distribution for the angular correlation function of B-modes $r=0.1$ with $\sigma_{\text{beam}}=2.7^{\circ}$ radian smoothing. The blue dashed line marks the $\Lambda$CDM prediction for the ensemble average.}\label{S12BB_r01}
\end{figure}

In order to present a meaningful correlation function and related statistics, we smooth the E- and B-mode power spectrum
with a $\sigma=2.7^{\circ}$ Gaussian beam (which corresponds to a $0.02$ radian beam). There are two benefits 
to this approach: it suppresses the 
$C_{\ell}^{BB}$ and $C_{\ell}^{EE}$ for $\ell\geq 50$
which ensures that the sum in Eq.~(\ref{Cxy}) converges, and it suppresses all pieces of the power spectrum that 
have contributions from lensing.  The former is necessary, since even for E- and B-mode
power spectra with perfect de-lensing, the sum in Eq.~(\ref{Cxy}) doesn't converge through $\ell_{\text{max}}=1500$. The latter is especially important since we wish to make statements about 
correlations of primordial E- and B-modes. Without smoothing we would need to de-lens all maps before calculating
statistics. At the smoothing level used for analysis here, lensing does not contribute to the calculated 
$S_{1/2}$ distribution. Therefore all results used here have been produced from power spectra that do not include
lensing effects.
 Figs.~\ref{CthetaBB_r01} and~\ref{CthetaEE_r01} show the resulting angular correlation function produced from the 
smoothed maps, and Figs.~\ref{S12EE_r01} and~\ref{S12BB_r01} show the distributions of $S_{1/2}$ statistics from 
simulations with $r=0.1$ (smaller values of $r$ will lead to an appropriate rescaling of the $\hat{B}\hat{B}$
distribution, but will leave other results unchanged).
 For a $\Lambda$CDM cosmology, the best-fit value of $S_{1/2}^{\hat{E}\hat{E}}$ is 
$1.86\times10^5\;\mu\mathrm{K}^4$
 and for $S_{1/2}^{\hat{B}\hat{B}}$ is $218.3\;\mu\mathrm{K}^4$.

A feature of the correlation functions of $\hat{E}({\bf \hat{n}})$ and $\hat{B}({\bf \hat{n}})$
being dominated by large multipoles, even for large angular scales. These functions are also not sensitive to the physics of
reionization, which make them a complimentary probe of correlation function suppression to the
$Q$ and $U$ correlations presented in the following section.

\section{$Q$ and $U$ Correlations}\label{QQUU}

The functions described in the section above may be undesirable in some cases, as they require taking
derivatives of observations. The $Q$ and $U$ correlation functions do not require 
derivaties, and have the added benefit that they are
 entirely dominated by the reionization bump terms with $\ell \leq 10$, avoiding the need for map smoothing
or concerns about contributions to the signal from lensing.

\begin{figure}
\includegraphics[scale=.6]{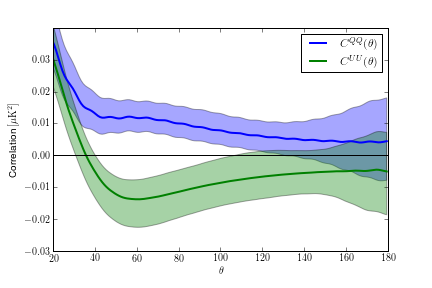}
\caption{Angular correlation function of $Q$ and $U$ polarizations with $r=0.1$. The shaded regions correspond
to the $68\%$ C.L. errors. The ranges include instrumental noise for a future generation PIXIE-like experiment 
and cosmic variance using Eq.~(\ref{pixie}).}\label{CthetaQU_r01}
\end{figure}

Fig.~\ref{CthetaQU_r01} shows the $QQ$ and $UU$ correlation functions for $r=0.1$ for $\Lambda$CDM. The shaded regions show the $68\%$ C.L. error 
regions for a PIXIE-like experiment plus cosmic variance calculated using Eq.~(\ref{pixie}).
There are distinct characteristics of the $QQ$ and $UU$ functions, namely that the $UU$ correlation is positive for a large 
range of angles while the $QQ$ function is negative for a large range of angles. Physical suppression
should drive both of these functions to zero. It could allow one to define additional 
measures of suppression of the correlation function beyond the standard $S_{1/2}$ statistic.

\begin{figure}
\includegraphics[scale=.6]{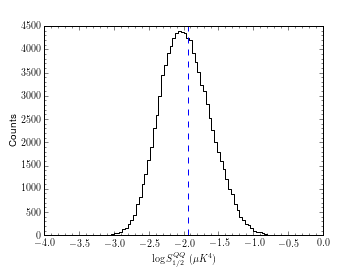}
\caption{$S_{1/2}$ distribution for $C^{QQ}(\theta)$ with $r=0.1$. The blue dashed line
shows the $\Lambda$CDM prediction for the ensemble average.}\label{CthetaQ_S12}
\end{figure}

\begin{figure}
\includegraphics[scale=.6]{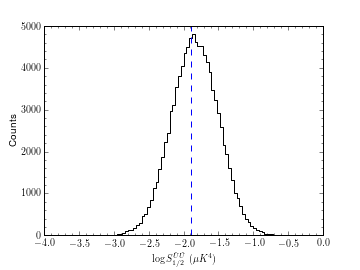}
\caption{$S_{1/2}$ distribution for $C^{QQ}(\theta)$ with $r=0.1$. The blue dashed line
shows the $\Lambda$CDM prediction for the ensemble average.}\label{CthetaU_S12}
\end{figure}

Figs.~\ref{CthetaQ_S12} and ~\ref{CthetaU_S12} show the $S_{1/2}$ distributions for both the QQ and UU correlation functions.
The $\Lambda$CDM value is shown with the blue dashed line. The expected $\Lambda$CDM value for $S_{1/2}^{QQ}$ is 
$0.0116\; \mu\mathrm{K}^4$ and for $S_{1/2}^{UU}$ is $0.0129\; \mu\mathrm{K}^4$.

In order to calculate $S_{1/2}$, the standard efficient methods defined in~\cite{Copi:2013zja} cannot be used.
Typically, Eq.~(\ref{Shalf}) is expanded to instead be a function of the $C_{\ell}$s and a coupling matrix using Eq.~(\ref{Cxy})
rather than calculating the integral of the square of $C(\theta)$ directly. Now, since Eq.~(\ref{QU}) is in terms of 
$G^{\pm}_{\ell}(\cos\theta)$ rather than $P_{\ell}(\cos\theta)$ as in Eq.~(\ref{Cxy}), the expressions for $S_{1/2}^{QQ}$ and 
$S_{1/2}^{UU}$ become more complicated. Appendix~\ref{dmat} describes a method that can be used to make the 
calculation more efficient by writing $G^{\pm}_{\ell}(\cos\theta)$ as functions of Wigner $d$-matrices.

The large-angle $Q$ and $U$ correlation functions being dominated by the reionization era, which is entirely inside the last scattering surface,
 give us a window into the nature of temperature suppression. The large-angle temperature correlation function has contributions from the last
scattering surface via the Sachs-Wolfe effect, and along the line of sight via the integrated Sachs-Wolfe effect.
The suppression of $C^{TT}(\theta)$, if caused by physics rather than a statistical fluke, could be due to features localized on the last
scattering surface alone or could include contributions from its interior.
If features inside the last scattering surface are suppressed, meaning suppression is a three-dimensional effect, this
will manifest as suppression in the $Q$ and $U$ correlation functions.

We have chosen to calculate the standard $S_{1/2}$ statistic, rather than generalizing to statistics at another
angle, $S(x)$, as defined in~\cite{Copi:2013zja}, since the reionization contribution is predominantly at $z=10$, which
is near enough to the surface of last scattering that the angular scale that features subtend are nearly that
of those at $z=1100$. Contributions from late-time reionization around $z=1$, which would skew the relevant angular scale,
 are subdominant since the amplitude 
of the polarization signal after $z_{\mathrm{reion}}$ falls off like $a^{-2}$. This leads to an overall 
drop-off in the correlation function of $a^{-4}$, meaning nearby effects are 100 times smaller than those
at $z=10$.

%%%%%%%%%%%%%%%%%%%%%%%%
\section{Conclusions}\label{conclusions}
%%%%%%%%%%%%%%%%%%%%%%%%

To address the lack of correlation in the temperature
power spectrum at large angles in particular, we need to move beyond temperature data alone. We show two viable methods for 
calculating correlation functions on the sky that arise from polarization and presented the distributions
for the corresponding statistics using constrained realizations for the E-mode 
contributions and the best-fit $\Lambda$CDM framework for B-mode realizations. A suppression in the primordial 
tensor or scalar fluctuations will affect the features of the two-point correlation function, 
meaning , local $C^{\hat{E}\hat{E}}(\theta)$ and $C^{\hat{B}\hat{B}}(\theta)$ as well as 
$C^{QQ}(\theta)$ and $C^{UU}(\theta)$, and their related statistical measures.
This would lend considerable weight to the 
argument that the lack of correlation seen in $C^{TT}(\theta)$ is due to primordial
physics, and is not just an anomalous statistical fluctuation of $\Lambda$CDM.   

We presented the distribution for an $S_{1/2}$ statistic for a $C^{\hat{B}\hat{B}}(\theta)$ from $\Lambda$CDM 
cosmology with $r=0.1$. If future limits on the value of $r$ are found to be significantly 
below this value, the results for $C^{\hat{B}\hat{B}}(\theta)$ will scale appropriately, 
wheras results for all other correlation functions will remain unchanged. For $C^{\hat{E}\hat{E}}(\theta)$, 
$C^{QQ}(\theta)$, and $C^{UU}(\theta)$, we considered constrained realizations, where $a_{\ell m}^E$ coefficients 
were related to $a_{\ell m}^T$ coefficients that match our power spectrum measurements and give values of 
$S_{1/2}^{TT}$ at least as small as we observe on the full- and cut-sky. We showed that for a $\Lambda$CDM
cosmology, the expected values of the statistics for Stokes parameter correlation functions are 
$S_{1/2}^{QQ} = 0.0116\;\mu\mathrm{K}^4$ and $S_{1/2}^{UU} = 0.0129\;\mu\mathrm{K}^4$, and the local E- and B-mode
expected values are $S_{1/2}^{\hat{E}\hat{E}} = 1.85\times 10^5\;\mu\mathrm{K}^4$ and 
$S_{1/2}^{\hat{B}\hat{B}} = 218.3\;\mu\mathrm{K}^4$. We chose to keep the previously defined $S_{1/2}$ for analysis
here, rather than generalizing to other angles than $\cos60^{\circ}=1/2$, as the dominant secondary
effect on polarization signals from epoch of reionzation is sufficiently close to the surface of last scattering to
not change the relevant angle of suppression significantly. Late-time reionization contributes
to the signal at a level 100 times smaller than the effect of reionization at $z=10$, so while those would
skew the relevant angular scales, they are subdominant.

Using a polarization error estimates for Planck, PIXIE and PRISM outlined in Table~\ref{params}, we calculated
the resulting $S_{1/2}$ statistics from a sky with exact suppression above $60^{\circ}$. These
values are presented in Table~\ref{tab:S12_exp}.  
We note that these levels are well below the $\Lambda$CDM predictions for all of the polarization
correlation functions presented here, and pixel noise for future experiments will not be a significant
source of error in identifying suppression. Measurement of large-angle polarization correlation functions
will have errors dominated by systematics rather than map pixel
noise for the foreseeable future.

Beyond being able to confirm that the suppression of temperature fluctuations is unlikely to be a statistical fluke, 
polarization correlation functions will add important new information. Because the local $\hat{E}$ and $\hat{B}$
correlation functions are dominated by large $\ell$ values, a suppression in all four correlation functions would strongly indicate that 
the suppression manifests itself physically in real-space at large angles. The 
$\hat{E}$ and $\hat{B}$ correlations give insight about suppression that is independent of
any effects of reionization which dominate the $Q$ and $U$ correlations.
 Also, foreground emission will contribute differently to the
various correlation functions.

Further, since the local $\hat{B}$ correlation is determined entirely by tensor fluctuations, a strong suppression
in that correlation function and not in others would show that the primordial suppression is predominantly in 
the tensor perturbations, while suppressions in local $\hat{E}$, $Q$ and $U$ but not in local $\hat{B}$ 
would suggest that the scalar perturbations are suppressed.

The distribution for $S_{1/2}$ statistics for each 
constrained correlation function was compared to the distribution
from $\Lambda$CDM alone. We found no significant difference between the two distributions and have presented only 
the constrained in this work. This means that polarization correlation functions provide
a largely independent probe of correlations compared to
the anomalous temperature correlation function. Future
high-sensitivity measurements of polarization over large 
fractions of the sky from envisioned experiments like PIXIE~\cite{Kogut:2011xw}
will differentiate primordial physics from a statistical fluke
as the origin of this anomaly. 

If the
suppressed temperature correlation is due to a statistical fluke, then
measurements of the polarization correlation function at large angular
scales is likely to give a much less suppressed signal. If, on the other
hand, the suppressed temperature correlation is due to some physical
mechanism, how well can polarization test this scenario? The answer
depends on the precise prediction of the suppression model. 
A Bayesian model comparison between a given model and the standard
cosmology will give a quantitative answer to this question. We are currently
investigating this possibility for a model with suppressed correlations
in the primordial gravitational potential perturbations. In general,
we expect suppressed primordial correlations will be evident in polarization
at least as much as in temperature, due to the lack of an integrated
Sachs-Wolfe contribution to the polarization perturbations. A
strong discrimination between a suppressed-correlation cosmology and
the standard cosmology is likely.

%%%%%%%%%%%%%%%%%%%%%%%%%%%%
\section{acknowledgements}
%%%%%%%%%%%%%%%%%%%%%%%%%%%%
The authors thank Sean Bryan, Ben Saliwanchik, and J.T. Sayre for useful conversations.
AY is supported by NASA NESSF Fellowship.  CJC, GDS and AY are supported
by a grant from the US DOE to the Particle Astrophysics Theory group at CWRU.
SA and AK are supported by NSF grant 1312380
through the Astrophysics Theory Program.

\appendix
\section{Correlation Functions and $S(x)$ Calculations for $QQ$ and $UU$ in Terms of Wigner d Matrices}\label{dmat}
The $G^{\pm}_{\ell}(\cos\theta)$ functions described in~\ref{QU} can be expressed in terms of reduced Wigner matrices.
This form may be useful for finding analytic expressions of $S_{1/2}^{QQ}$ and $S_{1/2}^{UU}$
which are easier to calculate numerically than performing the full integrals over $[C(\theta)]^2$. 
We can write the correlation functions for $Q$ and $U$ as
\begin{equation}
C^{QQ}(\theta) = \sum_{\ell} \frac{2\ell + 1}{8 \pi}\left(D_{\ell}^{+}(\cos\theta)C_{\ell}^{EE} + D_{\ell}^{-}(\cos\theta)C_{\ell}^{BB}\right)
\end{equation}
and
\begin{equation}
C^{UU}(\theta) = \sum_{\ell} \frac{2\ell + 1}{8 \pi}\left( D_{\ell}^{-}(\cos\theta)C_{\ell}^{EE} + D_{\ell}^{+}(\cos\theta)C_{\ell}^{BB}\right)\;,
\end{equation}
where we have assumed parity invariance and used
\begin{equation}
D^{\pm}_{\ell}(\cos\theta)\equiv\left[ d_{2,2}^{\ell}(\theta) \pm d_{2,-2}^{\ell}(\theta)\right]
\end{equation}
from~\cite{Chon:2003gx}, where $d_{i,j}^{\ell}(\theta)$ are the reduced Wigner rotation matrices.

This form of the correlation functions would lead to a method of calculating $S_{1/2}$ most similar
to that defined in~\cite{Copi:2013zja} by using properties of d-matrix integrals and recursion relations.

The general form of the S statistic is defined as
\begin{equation}
S(x) \equiv \int_{-1}^{x}\left[ C^{XX}(x)\right]^2\; \mathrm{d}x, 
\end{equation}
where $x=\cos\theta$ and $S_{1/2}=S(1/2)$. We define
\begin{equation}\label{Iint}
I_{\ell\ell^{\prime}}^{\pm\pm}\equiv\int_{-1}^{x}d_{2,\pm 2}^{\ell}(x)d_{2,\pm 2}^{\ell^{\prime}}(x)\;\mathrm{d}x,
\end{equation}
which we use to calculate $S(x)$ using properties of the reduced Wigner matrices.

Important properties of the $I_{\ell\ell^{\prime}}^{\pm\pm}$ matrices are that $I_{\ell\ell^{\prime}}^{+-}(x) = I_{\ell\ell^{\prime}}^{-+}(x)$ and
\begin{equation}
I_{\ell\ell^{\prime}}^{--}(x) = (-1)^{\ell+\ell^{\prime}} \left[\frac{2}{2\ell +1}\delta_{\ell\ell^{\prime}}
-  I_{\ell\ell^{\prime}}^{++}(-x)\right], 
\end{equation}
so all required quantities can be constructed from calculating $I_{\ell\ell^{\prime}}^{+-}(x)$ 
and $I_{\ell\ell^{\prime}}^{++}(x)$ only. To compute these, we use the relation betwen reduced Wigner matrices
and the Clebsch-Gordan coefficients~\cite{angmom}, $C^{jm+m^{\prime}}_{\ell m \ell^{\prime}m^{\prime}}$:
\begin{align}\label{CG}
& d_{m_1,m_2}^{\ell}(x)d_{m^{\prime}_1,m^{\prime}_2}^{\ell^{\prime}}(x) = \notag\\
& \sum^{\ell+\ell^{\prime}}_{j=|\ell-\ell^{\prime}|}
C^{jm_1+m_1^{\prime}}_{\ell m_1 \ell^{\prime}m^{\prime}_1}C^{jm_2+m_2^{\prime}}_{\ell m_2 \ell^{\prime}m^{\prime}_2}
d^j_{m_1+m^{\prime}_1\;m_2+m^{\prime}_2}(x).
\end{align}
Combining Eq.~(\ref{Iint}) with Eq.~(\ref{CG}) and exploiting properties of the Clebsch-Gordan coefficients, we find:
\begin{equation}
I_{\ell\ell^{\prime}}^{++}(x) = \sum^{\ell+\ell^{\prime}}_{j=\mathrm{max}(|\ell-\ell^{\prime}|,4)}
\left[C_{\ell 2 \ell^{\prime} 2}^{j4}\right]^2\int_{-1}^{x} d_{44}^{j}(x)\; \mathrm{d}x
\end{equation}
and
\begin{equation}
I_{\ell\ell^{\prime}}^{+-}(x) = \sum^{\ell+\ell^{\prime}}_{j=\mathrm{max}(|\ell-\ell^{\prime}|,4)}
C_{\ell 2 \ell^{\prime} 2}^{j4}C_{\ell 2 \ell^{\prime} -2}^{j0}\int_{-1}^{x} d_{40}^{j}(x)\; \mathrm{d}x.
\end{equation}
The integrals have analytic solutions, which make computation of the $I_{\ell\ell^{\prime}}$ matrices
more efficient.
The integral over $d_{40}^j(x)$ is the simpler of the two cases:
\begin{equation}
i_j^{(0)}(x) \equiv \int_{-1}^x d_{40}^j(x) \,\mathrm{d}x.
\end{equation}
This integral can be performed by noting that
\begin{equation}
d_{m 0}^j(\beta) = \sqrt{\frac{4\pi}{2j+1}} Y_{j m}^*(\beta,0) = \sqrt{\frac{(j-m)!}{(j+m)!}} P_j^m(\cos\beta).
\end{equation}
Using the Rodrigues formula,
\begin{equation}
P_j^m(x) = (-1)^m (1-x^2)^{m/2} \frac{\mathrm{d}^m}{\mathrm{d}x^m} P_j(x),
\end{equation}
and integrating by parts, we can show that
\begin{align}
i_j^{(0)}(x) &= \sqrt{\frac{(j-4)!}{(j+4)!}} \int_{-1}^x (1-x^2)^2 \frac{\mathrm{d}^4}{\mathrm{d}x^4} P_j(x) \,\mathrm{d}x \notag \\
 & = \sqrt{\frac{(j-4)!}{(j+4)!}}\left( -\sqrt{1-x^2} P_j^3(x) \right. \notag \\
 & + 4x P_j^2(x) - 4(1-3x^2) \frac{\mathrm{d}P_j(x)}{\mathrm{d}x} \notag\\
 & + 4(-1)^j j(j+1) - 24 x P_j(x) - 24(-1)^j \notag \\
 & \left. + \frac{24}{2j+1} \left[ P_{j+1}(x) - P_{j-1}(x) \right]\right).
\end{align}

The integral over $d_{44}^j(x)$ is more conveniently done as integrals over angles:
\begin{equation}
i_j^{(4)}(x) \equiv \int_{\cos^{-1}(x)}^\pi d_{44}^j(\beta) \sin\beta \,\mathrm{d}\beta.
\end{equation}
Using the relation
\begin{align}
\sin\beta \, d_{44}^j(\beta) &= \frac{\sqrt{(j^2-4^2)(j+3)(j+4)}}{j(2j+1)} d_{43}^{j-1}(\beta) \notag \\
&-\frac{4\sqrt{(j-3)(j+4)}}{j(j+1)} d_{43}^j(\beta) \notag \\
&- \frac{\sqrt{[(j+1)^2-4^2](j-3)(j+2)}}{(j+1)(2j+1)} d_{43}^{j+1}(\beta),
\end{align}
along with properties of integrals over $d_{43}^j(\beta)$, $d_{42}^j(\beta)$, $d_{41}^j(\beta)$,
and $d_{40}^j(\beta) / \sin(\beta)$, 
the integral becomes
\begin{align}
i_j^{(4)}(x) &= \frac{\sqrt{(j^2-4^2)(j+3)(j+4)}}{j(2j+1)} k_{j-1}(x) \notag\\
&-\frac{4\sqrt{(j-3)(j+4)}}{j(j+1)} k_j(x) \notag \\
&- \frac{\sqrt{[(j+1)^2-4^2](j-3)(j+2)}}{(j+1)(2j+1)} k_{j+1}(x),
\end{align}
where
\begin{align}
k_j(x) &= \sqrt{\frac{(j+2)(j-1)}{(j-2)(j+3)}} \ell_j(x) \notag\\
&\quad\quad + \frac{2}{\sqrt{(j-2)(j+3)}} d_{42}^j(x), \\
\ell_j(x) &= \frac{1}{\sqrt{j(j+1)}} d_{40}^j(x) - \frac{4}{\sqrt{j(j+1)}} m_j(x),
\end{align}
and
\begin{align}
m_j(x) &= \sqrt{\frac{(j-4)!}{(j+4)!}} \left( -\frac{1}{\sqrt{1-x^2}} P_j^3(x) \right. \notag \\
& + \frac{2x}{1-x^2} P_j^2(x) + \frac{(-1)^j}{4} \frac{(j+2)!}{(j-2)!} \notag \\
&\left. - 2 \frac{\mathrm{d}}{\mathrm{d}x} P_j(x) - (-1)^j j(j+1) \right).
\end{align}
Recursion relations were used to calculate the Clebsch-Gordan coefficients from
Eqs. (8.5:3), (8.5:8), and (8.6:27) in~\cite{angmom}.

We compute the $I_{\ell\ell^{'}}^{(i)}$ matrices in Eq.~(\ref{IQU}) directly from these forms, as
\begin{eqnarray}
I_{\ell\ell^{'}}^{(1)}(x) &=& I_{\ell\ell^{\prime}}^{++}(x) + 2I_{\ell\ell^{\prime}}^{+-}(x) +
I_{\ell\ell^{\prime}}^{--}(x) \notag\\
I_{\ell\ell^{'}}^{(2)}(x) &=& I_{\ell\ell^{\prime}}^{++}(x) + I_{\ell\ell^{\prime}}^{--}(x)\notag\\
I_{\ell\ell^{'}}^{(3)}(x) &=&I_{\ell\ell^{\prime}}^{++}(x) - 2I_{\ell\ell^{\prime}}^{+-}(x) +
I_{\ell\ell^{\prime}}^{--}(x).
\end{eqnarray}

\end{document}